\documentclass[peerreview,onecolumn,11pt,draftclsnofoot]{IEEEtran}

\usepackage{amsmath}
\usepackage{amsmath,amsthm,amssymb}
\usepackage{color,soul}
\usepackage{amsfonts}
\usepackage{epsfig}
\usepackage{amssymb}
\usepackage{cite}
\usepackage{subfigure}
\usepackage{multirow}
\usepackage{rotating}
\usepackage{graphicx}
\usepackage{tabularx}
\usepackage{array}
\usepackage{amsfonts}
\usepackage{graphicx}  
\usepackage{graphics}  
\usepackage{epsfig}
\usepackage[all]{xy}
\usepackage{tabularx}
\usepackage{amsmath}
\usepackage{amsfonts}
\usepackage{epstopdf}
\usepackage{amssymb}
\usepackage{cite}
\usepackage{subfigure}
\usepackage{multirow}
\usepackage{rotating}
\usepackage{graphicx}
\usepackage{tabularx}
\usepackage{array}
\usepackage{setspace}
\usepackage{morefloats}
\usepackage{float}

\makeatletter
\newcommand*{\rom}[1]{\expandafter\@slowromancap\romannumeral #1@}
\makeatother

\usepackage{amsmath,amssymb}
\usepackage{amsmath}
\usepackage{algorithm}
\usepackage[noend]{algpseudocode}

\usepackage{float}

\usepackage{algorithm,algpseudocode}
\makeatletter
\newcommand{\algmargin}{\the\ALG@thistlm}
\makeatother
\newlength{\whilewidth}
\algdef{SE}[parWHILE]{parWhile}{EndparWhile}[1]
  {\parbox[t]{\dimexpr\linewidth-\algmargin}{%
     \hangindent\whilewidth\strut\algorithmicwhile\ #1\ \algorithmicdo\strut}}{\algorithmicend\ \algorithmicwhile}%
\algnewcommand{\parState}[1]{\State%
  \parbox[t]{\dimexpr\linewidth-\algmargin}{\strut #1\strut}}

\newtheorem{theorem}{Theorem}

\makeatletter
\def \BState{\State \hskip - \ALG@thistlm}
\makeatother
\begin{document}

\title{\LARGE{Cooperative Energy Efficient Power Allocation Algorithm for Downlink Massive MIMO}
\author{Saeed~Sadeghi~Vilni}
\thanks{The author is with the Department of Electrical and Computer Engineering, Tarbiat Modares University, P.~O.~Box 14115-194, Tehran, Iran. Corresponding author is S. S. Vilni (email: saeed.sadeghi@modares.ac.ir).}}

\maketitle

\begin{abstract}

Massive multiple input multiple output (MIMO) is proposed to increase network capacity and reduce power consumption in next generation of wireless networks. In this paper, power allocation in the massive multiple input multiple output (MIMO) cellular networks is studied. Considering circuit power consumption, energy efficiency is utilized to evaluate the network performance, since circuit power consumption increases by increasing both the number of antenna in transmitter and the number of network users. To optimize transmit power of the network, an energy efficient optimization problem is considered under both maximum transmit power and quality of service (QoS) constraints. Where the users cooperate with the base station (BS) to minimize network power usage. To solve the problem, an energy efficient power allocation algorithm is proposed. Convergence of the proposed algorithm is shown numerically, in addition, the appropriate number of transmit antenna and users are obtained. Finally, pilot contamination effect is evaluated, where it is shown energy efficiency decreases dramatically.
\end{abstract}

\begin{IEEEkeywords}
Massive MIMO, power allocation, energy efficiency, cooperation, pilot contamination.
\end{IEEEkeywords}

\section{Introduction}\label{Sect1}

Cellular network traffic is increased greatly todays, due to rapid growth of smart terminal usage and high data rate requirements services \cite{survey}-\cite{survey2}. In order to increase the spectral and energy efficiency up to around ten times massive multiple input multiple output (MIMO) is proposed in the next generation of wireless networks. Massive MIMO means using large number of transmit antenna (about one hundred antenna) and can enhance spectral efficiency and reliability \cite{massive}-\cite{massive2}.

 With a large number of transmit antenna, massive MIMO can focus energy at desired point and increase spectral efficiency but circuit power consumption, which connected to the number of antenna, also increased. 
 To achieve maximum energy efficiency, unlimited number of transmit antenna is optimal if circuit power consumption is not considered \cite{CP}. Hence to evaluate network energy efficiency, which is throughput to power consumption ratio, circuit power consumption should be taken into account.
 
Energy efficient resource allocation is one of open problems in wireless system design, wich is considered in \cite{ee41}-\cite{ee10}. In \cite{ee41}-\cite{ee43} authors try to maximize the energy efficiency but circuit power consumption is ignored; Also in \cite{ee51} authors assume a constant circuit power consumption.

From another point of view, QoS plays an important role in wireless system resource allocation; for example, satisfying QoS constraint in addition to providing a roughly fair network leads a practical network. No QoS constraint is considered in \cite{ee43}-\cite{ee72}. In \cite{ee41} and \cite{ee71}-\cite{ee9}, authors assume that the number of transmit antenna is very greater than users, which is impractical assumption. In this paper, we try to make this assumption practical, we consider any number of transmit antenna and users. In \cite{ee10} authors consider imperfect channel state information (CSI), but in system design, they utilize estimated channel and real channel together; which appears they have a design error. 

\begin{figure}[H]
\center
\includegraphics[scale = 0.23]{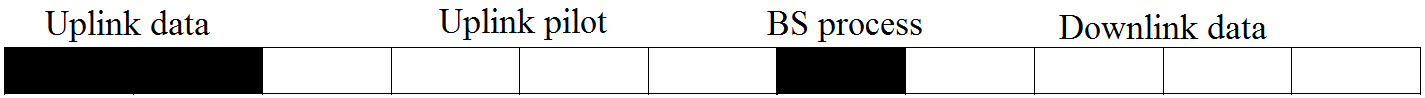}
\caption{TDD protocol to acquire CSI}
\label{Fig1}
\end{figure}

To obtain channel state information (CSI), The base stations (BSs) use time division duplexing (TDD) protocol as shown in Fig. \ref{Fig1}. As the protocol shows, users send a pilot sequence after their data to the BS. The BS estimates the users' channel using the received pilot and sends the channel data back to users. Note that in spite of the pilots are sent in the same bandwidth, but they must be orthogonal to data; hence this is not the perfect CSI. Due to resource limitation, it is not possible for all users in all cells to have orthogonal pilots, therefore each pilot is reused  by users of the neighbor cells, this creates pilot contamination. Pilot contamination leads to imprecisely channel estimation which reduces system performance \cite{pcon}. Pilot contamination in \cite{ee51} and \cite{ee71}-\cite{ee9} is not considered too.

 In this paper, we investigate energy efficient power allocation in a cellular network with large number of transmit antenna (massive MIMO), that users cooperate with their BS. We consider energy efficiency as users' utility where overall transmit power in a cell is limited. In addition, a minimum data rate should be provided for each user. We utilize the fractional programming technique to solve the problem, since the energy efficiency objective function is a fractional function.  We assume cooperation between users and their BS, and found a closed form for optimal transmit power of users. The algorithm of cooperative energy efficient power allocation is given in Algorithm 1. We extend the problem to  multi cell case under pilot contamination. Finally, the convergence of the proposed algorithm is investigated by simulation. In addition, we explore pilot contamination effect on energy efficiency.

The remainder of this paper is organized as follows. In Section II the system model is presented. In Section III the energy efficient optimization problem is formulated. The solution for the problem is presented in Section IV. In section V our scenario extended to multi cell case. Numerical results are described in Section VI followed by conclusion in Section VII.

Notation: Vectors and matrices are denoted by boldface small and big letters, respectively. The superscript $H$ stands for conjugate transpose. Euclidean norm with $\|.\|$ is denoted and ${\bf I}_K$ shows an Identity $K\times K $ matrix. $\mathcal{N}(0,\sigma^2)$ denotes a Gaussian probability density function with zero mean and variance $\sigma^2$.

\section{System Model}\label{sect2}

Consider a downlink single cell network including a BS equipped with {\it M} antennas and {\it K} single antenna users which are distributed in the cell uniformly. Assume flat fading channels between BS and users with perfect channel state information (CSI) and all users get service on a single frequency.

Channel vector between BS and user {\it i} is modeled by $\mathrm{\bf g}_{i}= \sqrt{\beta_i} \times \mathrm{\bf h}_i\in  \mathbb{C}^{ M \times 1}$ where $\mathrm{\bf h}_i$ shows fast fading and $\beta_i$ is large scale fading that consists of path loss and shadow fading. Fast fading has a distribution as $\mathcal{CN}(0,1)$ and $\beta_i = \psi(\frac{d_0}{d_i})^v$ where $\psi$ shows shadow fading with the distribution $10\log_{10}^{\psi}\sim \mathcal{N}(0, \sigma_{sh}^{2})$, $d_0$ is minimum distance from users to the BS, and $d_i$ is distance from user {\it i} to the BS and $v$ is the path loss exponent.

The received signal by user $i$ can be expressed as
\begin{equation}
y_i = \sum^{K}_{k = 1} \sqrt{p_k}{\bf g}_i{\bf w}_kx_k + n_i
\label{eq1:received-signal}
\end{equation}
Where $p_k$ is transmit power of user $k$ which is an element of transmit power vector ${\bf p} = [p_1 ... p_k ... p_K]$, $x_k$ is transmission symbol for user $k$ with $\|x_k\|^2 = 1$ which is assumed that transmission symbols are uncorrelated among different users, $n_i$ is additive noise with $\mathcal{CN}(0,\sigma^2)$ distribution and ${\bf w}_l \in \mathbb{C}^{M\times 1}$ is beamforming vector for $l$th user, by maximum ratio transmission (MRT) beamforming we have
\begin{equation}
{\bf w}_k = \frac{{\bf g}_k^{H}}{\|{\bf g}_k\|}
\label{eq2:beamforming-vector}
\end{equation}
The received signal by user $i$ can be written as
\begin{equation}\label{eq3:seprat-signal-interference}
y_i = \sqrt{p_i}{\bf g}_i{\bf w}_ix_i + \sum^{K}_{k = 1, k \neq i} \sqrt{p_k}{\bf g}_i{\bf w}_kx_k + n_i
\end{equation}
Equation \eqref{eq3:seprat-signal-interference} is the summation of three parts: desired signal, interference and noise. According to \eqref{eq3:seprat-signal-interference}, SINR of user $i$ can be expressed as following.
\begin{equation}
sinr_i(p_i) = \frac{p_i\|{\bf g}_i{\bf w}_i\|^2}{\sum^{K}_{k = 1, k \neq i} p_k\|{\bf g}_i{\bf w}_k\|^2 + \sigma^2}
\end{equation}
Where $\sigma^2$ is noise power on the channel of user $i$. The instantaneous data rate for user $i$  can be written as following
\begin{equation}
r_i(p_i) = B\log(1 + \Gamma_isinr_i(p_i))
\end{equation}
Where $B$ is bandwidth and $\Gamma_i$ is SINR gap between Shannon channel capacity and a practical modulation and coding scheme. This SINR gap is equal to $\frac{-2}{3\ln (5e_i)}$ that $e_k$ is target bit error rate \cite{shgap}.
\section{Problem Statement}
\subsection{Energy Efficiency}
Power consumption in the network includes transmit power and power consumed in the circuits. Circuit power consumption in BS has two parts: constant part $P^\mathrm{fix}$ involved site cooling, transmitting filter, converter and local oscillator; and variable part, $P^\mathrm{a}$, is the power to run an antenna. If circuit power consumption of each user is displayed with $P^{\mathrm{ue}}$, the power consumption $\mathcal P$ can be defined as 
\begin{equation}
\mathcal P({\bf p}) = \sum^{K}_{k = 1}p_l + (M\times P^\mathrm{a})+ P^\mathrm{fix} + (K\times P^{\mathrm{ue}})
\end{equation}
According to \cite{ee1} and \cite{ee2} the energy efficiency $ \eta $ in communication system calculated in bit/Joule and is the ratio of total rate to power consumption which can be formulated as the following
\begin{equation}\label{EE}
\eta({\bf p}) = \frac{\sum^{K}_{k = 1}r_l(\bf p)}{\mathcal P(\bf p)}
\end{equation}
\subsection{Optimization Problem}

According to above discussion, the goal of the network is to maximize the energy efficiency \eqref{EE} under maximum power and minimum data rate constraints, i.e. following problem
\begin{equation}\label{opt1}
\begin{array}{rlllll}
\displaystyle {\max_{\bf p} }& \multicolumn{1}{l}{\eta({\bf p})} \\
\textrm{s.t.} &  C1: \sum_{k=1}^{K}p_{k}\leq P^{max}\\
&\displaystyle  C2: r_{k}({\bf p})\geq R^{min}~\forall k
\end{array}
\end{equation}
Therefore, the network should allocate transmit power to each users such that maximize the energy efficiency. In \eqref{opt1}, constraint $C1$ is the power constraint where $P^\mathrm {max}$ is total maximum transmit power, and constraint $C2$ shows minimum data rate $R^{min}$ that must be provided for each user.

\section{Solving the Problem}
There are two main challenges in \eqref{opt1}: the problem has fractional form and \eqref{EE} is not concave. To solve the problem \eqref{opt1}, we transform the the objective function of \eqref{opt1} to a subtractive form by the following Theorem \cite{dinkel}.
\begin{theorem}\label{t1}
The maximum energy efficiency $\eta ^*$ is achieved in (\ref{opt1}) \textit{if and only if} $\displaystyle \max_{\bf p}$ $\sum^{K}_{k = 1}r_k({\bf p})-\eta ^*\mathcal{P({\bf p})}=\sum^{K}_{k = 1}r_k({\bf p^*})-\eta ^*\mathcal{P({\bf p^*})}=0$ for $\sum^{K}_{k = 1}r_k({\bf p})\geq0$ and $\mathcal{P({\bf p})} > 0$.
\end{theorem}

 According to Theorem \ref{t1}, the problem \eqref{opt1} is transformed to following problem.
\begin{equation}\label{opt2}
\begin{array}{rlllll}
\displaystyle {\max_{\bf p} }& \multicolumn{1}{l}{\sum^{K}_{k = 1}r_k({\bf p})-\eta\mathcal{P({\bf p})}} \\
\textrm{s.t.} &  C1: \sum_{k=1}^{K}p_{k}\leq P^{max}\\
&\displaystyle  C2: r_{k}({\bf p})\geq R^{min}~\forall k
\end{array}
\end{equation}
To solve \eqref{opt1}, we solve (\ref{opt2}) iteratively. We propose to solve problem (\ref{opt2}) with a specific value of energy efficiency $\eta$ and obtain users transmit power ${\bf {p}}_{t_1}$ for iteration $t_1$; and  then compute $A = \sum^{K}_{k = 1}r_k({{\bf p}_{t_1}})-\eta\mathcal{P}({{\bf p}_{t_1}})$. If $A$ is zero, $\eta$ is optimal energy efficiency and ${{\bf p}_{t_1}}$ is optimal transmit power; otherwise energy efficiency with ${{\bf p}_{t_1}}$ is computed and then ${{\bf p}_{t_1}}$ is used in (\ref{opt2}) and solve the problem again until $A$ goes to zero.

To address non-concavity of the objective function of \eqref{opt1}, we propose to relax the objective function by the following lower bound \cite{scale}.
\begin{equation}\label{sca1}
\log{(1 + \Gamma_i sinr_i(p_i))}\geq a_i\log{ (\Gamma_i sinr_i(p_i))} + b_i
\end{equation}
Where $a_i$ and $b_i$ are auxiliary multiplicative and additive variables for user $i$. Equation \eqref{sca1} means utilizing  $\hat r_i = a_i\log{ (\Gamma_i sinr_i(p_i))} + b_i$ instead of $r_i = \log{(1 + \Gamma_i sinr_i(p_i))} $ as the objective function. The variable of problem $\mathbf{p}$ is changed to $\bf \hat  p = \ln \bf p$. Therefore we have the following problem.
 \begin{equation}\label{opt3}
\begin{array}{rlllll}
\displaystyle {\max_{\bf \hat p} }& \multicolumn{1}{l}{\sum^{K}_{l = 1}\hat r_l({\bf \hat p})-\eta\mathcal{P({\bf \hat p})}} \\
\textrm{s.t.} &  C1: \sum_{k=1}^{K}e^{\hat p_{k}}\leq P^{max}\\
&\displaystyle  C2: \hat r_{k}({\bf \hat p})\geq R^{min}~\forall k
\end{array}
\end{equation}

Since log-sum-exp is concave \cite{boid}, The problem (\ref{opt3}) is a convex optimization problem. In order to solve (\ref{opt2}), we solve (\ref{opt3}) iteratively until transmit power converged; also $a_i$ and $b_i$ are updated in each iteration by obtained transmit power from current iteration with following equation.
\begin{subequations}
\begin{equation}
a_i = \frac{sinr_i(\hat p_i)}{1+sinr_i(\hat p_i)}
 \end{equation}
 
 \begin{equation}
 b_i = \log(1+sinr_i(\hat p_i))- a_i\log(sinr_i(\hat p_i))
 \end{equation}
 
\end{subequations}
Since the problem \eqref{opt3} is a convex optimization problem, dual  Lagrangian function is utilized to solve that. Let $\lambda$ and $\phi$ Lagrange multiplier corresponding to maximum transmit power and minimum data rate constraint, the dual Lagrangian function of (\ref{opt3}) can be written as following \cite{boid}.
\begin{equation}
\begin{array}{l}
\mathcal{L}(\bf \hat p, \lambda, \phi) \\
=\sum^{K}_{k = 1}\hat r_k({\bf \hat p})-\eta\mathcal{P({\bf \hat p})}\\
-\phi (\sum_{k=1}^{K}e^{\hat p_{k}} - P^{\mathrm{max}})\\
+ \sum_{k=1}^{K}\lambda_l(R^{\mathrm{min}}-r_k)
\end{array}
\end{equation}

Optimal transmit power can be obtained by the Karush Kuhn Tucker (KKT) conditions as following for user $i$.
\begin{equation}\label{L1}
\frac{\partial{\mathcal{L}(\bf \hat p, \lambda, \phi)}}{\partial{p_i}} = 0
\end{equation}

According to (\ref{L1}), optimal transmit power for user $i$ can be obtained as following.
\begin{equation}\label{optp}
e^{\hat p_i} = p_i = [\frac{ (\lambda_i + 1)\frac{Ba_i}{ln2}}{\frac{B}{ln2}\sum_{k = 1, k \neq i}^{K}\frac{(\lambda_k +1)a_k(\| w_ig_k\|^2)}{I_k}+(\eta+\phi)}]^+
\end{equation}
Where $I_k$ is interference plus noise suffered by user $k$. In this paper cooperating of users with BS is considered such as they measure interference $I_k$ and feed back to BS. $I_k$ can be calculated as
\begin{equation}
I_k = \sum_{u = 1, u\neq k}^{K}{e^{\hat p_u(\|w_uh_k\|^2)}} + \sigma^2
\end{equation}

By the subgradient method the Lagrange multipliers can be updated as following.
\begin{equation}\label{phi}
\phi (t_3 + 1) = [\phi(t_3) + \gamma_{\phi}(\sum_{k=1}^{K}e^{\hat p_{k}} - P^{\mathrm{max}})]^+
\end{equation} 
\begin{equation}\label{lambda}
\lambda_k(t_3 + 1) = [\lambda_l(t_3) + \gamma_{\lambda}(R^{\mathrm{min}}-r_k)]^+
\end{equation}
Where $t_3$ is iteration index for Lagrange multipliers update also $\gamma_{\phi}$ and $\gamma_{\lambda}$ are step size for $\phi$ and $\lambda$ respectively. Algorithm for cooperative energy efficient power allocation is presented in Algorithm 1.

\section{Extension to Multi Cell Case}

In this section, the single cell scenario in Section \ref{sect2} is extended to multi cell case, where each cell has $K$ users and all users operate in a single frequency. Pilot contamination and estimate channel with least square (LS) method is considered in multi cell case as in \cite{ls}. As previous section optimal transmit power for user $m$ in cell $j$ can be obtained as following
\begin{equation} \label{opt.P4}
p_{jm} = [\frac{(\lambda_{jm}+1)\frac{B\alpha_{jm}}{\ln2}}{\frac{B}{\ln2}\sum_{l=1}^{L}\sum_{k=1,k\neq m}^{K}\frac{z_{lk}}{I_{lk}}-(\eta +\phi_j)}]^+
\end{equation}
Where $z_{lk}= \|w_{jm}\hat g_{jlk}\|^2 \alpha_{lk} (\lambda_{lk}+1)$,  $\hat{g}_{jlk}$ shows estimated channel between user $k$ in cell $l$ and $j$th BS, $I_{lk}$ is interference plus noise that measured by users and send for corresponding BSs, $\alpha_{lk}$ is SCA multiplicative variable, $\lambda_{lk}$ is Lagrange multiplier corresponding to minimum data rate constraint and $w_{lk}$ is beamforming vector for user $k$ in cell $l$, respectively. Also, $L$ shows the number of cells, $\phi_{j}$ is the Lagrange multiplier corresponds to maximum transmit power in cell $j$ and $\eta$ is network energy efficiency. The method to solve the optimization problem is given in the appendix A.

\begin{algorithm} 
\caption{Energy Efficient Cooperative Power Allocation}\label{A1}
\begin{algorithmic}[1]
\State Initialize convergence tolerance $\epsilon$ and initialize arbitrary $\eta(t_1)=0$, ${con} = 0$ and Dinkelbach algorithm iteration index $t_1=1$
\Repeat
\State initialize with a feasible ${{\bf \hat p}_{t_2}}$, Set $a_{l}(t_2)=1$,
\parState{$b_{l}(t_2)=0$ and SCA iteration index  $t_{2}=1$}
\Repeat
\parState  {initialize arbitrary $\lambda_{l}(t_3)$, $\phi(t_3)$, $\gamma_\lambda$, $\gamma_\phi$ and {Lagrangian {multipliers iteration index $t_3 = 1$}}}
\Repeat
\parState{Compute $I_{l}$ by user and feed back to {\bf BS} }
\parState{Compute the optimal power transmit ${\bf \hat p}_{t_2+1}$ according to (\ref{optp})}
\State Update $\phi$ and $\lambda_{l}$ according to (\ref{phi}) and (\ref{lambda})
\State{$t_3 = t_3 + 1$}
\Until {Convergence of $\lambda_{l}$ and $\phi$}
\parState{Compute $sinr_{l}({\bf \hat p}_{t_2+1})$ and update $a_{l}(t_2+1)$ and $b_{l}(t_2+1)$}
\State {$t_2 = t_2 + 1$}
\Until {convergence of $\bf \hat p$ }
\If {$\sum^{K}_{l = 1}\hat r_l({{\bf \hat P}(t_2+1)})-\eta(t_1)\mathcal{P}({{\bf \hat p}_{t_2 + 1}})\leq \epsilon $}
\State Set $con = 1$
\State {\bf Return} $\eta(t_1)$ as optimal energy efficiency and $\hat {\bf p}$
\parState{ as optimal transmit power}
\Else
\State Set $con = 0$
\State $t_1 = t_1 + 1$
\State {\bf Return} $\eta({t_1 + 1}) = \frac{\sum^{K}_{l = 1}r_l(\hat {\bf p}_{t_2})}{\mathcal P(\hat {\bf p}_{t_2})}$
\EndIf
\Until {${con} = 1$}
\end{algorithmic}
\end{algorithm}
\begin{table}[h]\label{j1}
\caption{Simulation Parameters} 
\centering          
\begin{tabular}{|l | c | r|} 
\hline     
Parameter & Description & Value\\
\hline
$\sigma^2$ & Noise power & -174dBm\\
\hline
$\sigma^2_{sh}$&Shadow fading variance&8dB\\
\hline
$P^{\mathrm{max}}$& Total maximum transmit power& 1W\\
\hline
$P^{\mathrm{a}}$& Power consumption per antenna& 1W\\
\hline
$P^{\mathrm{ue}}$& Power consumption per antenna& 0.1W\\
\hline
$P^{\mathrm{fix}}$& Fixed power consumption in BS& 20W\\
\hline
$R^{\mathrm{min}}$&Minimum requirement rate for users& 14Kbps\\
\hline
$B$&Bandwidth & 10KHz\\
\hline
$e_k$&Target bit error rate & 10e-3\\
\hline
-&Cell radius & 500m\\
\hline
$d_0$& Minimum distance & 50\\
\hline
$\alpha$& Path loss exponent&3\\
\hline

\end{tabular}
\end{table}

\section{Analysis and Simulation Results}\label{sect4}

In this section, we evaluate cooperative energy efficient power allocation in algorithm \ref{A1}. Simulation parameters are given in Table I.
First, the convergence of the Algorithm 1 is shown.
 Fig. 2 shows the convergence of algorithm \ref{A1} when number of antennas $M = 100$ and number of users $K = 5$, as it can be seen, the algorithm converges in three iterations.

In fig. 3 energy efficiency versus number of transmit antenna $M$ is shown for $K = 5$; which is a concave curve. When the number of transmit antennas increase, the performance of beamforming gets better; therefor SINR and data rate grow. On the other hand, circuit power consumption increased when $M$ increased. In figure 2, first, increasing of data rate overcome circuit power consumption but increasing number of antenna leads circuit power consumption overcome data rate increases.

To consider the effect the number of users, energy efficiency in terms of the number of users is depicted in fig. 4. It can be seen that firstly, energy efficiency increased but by increasing the number of users near to number of antenna energy efficiency start decreasing. this is due to inter user interference increasing.

Fig. 5 shows average transmit power for one user versus $M$. It shows that transmit power decreases by increasing the number of transmit antennas which is expected.
 
\begin{figure}
\center
\includegraphics[scale=0.58]{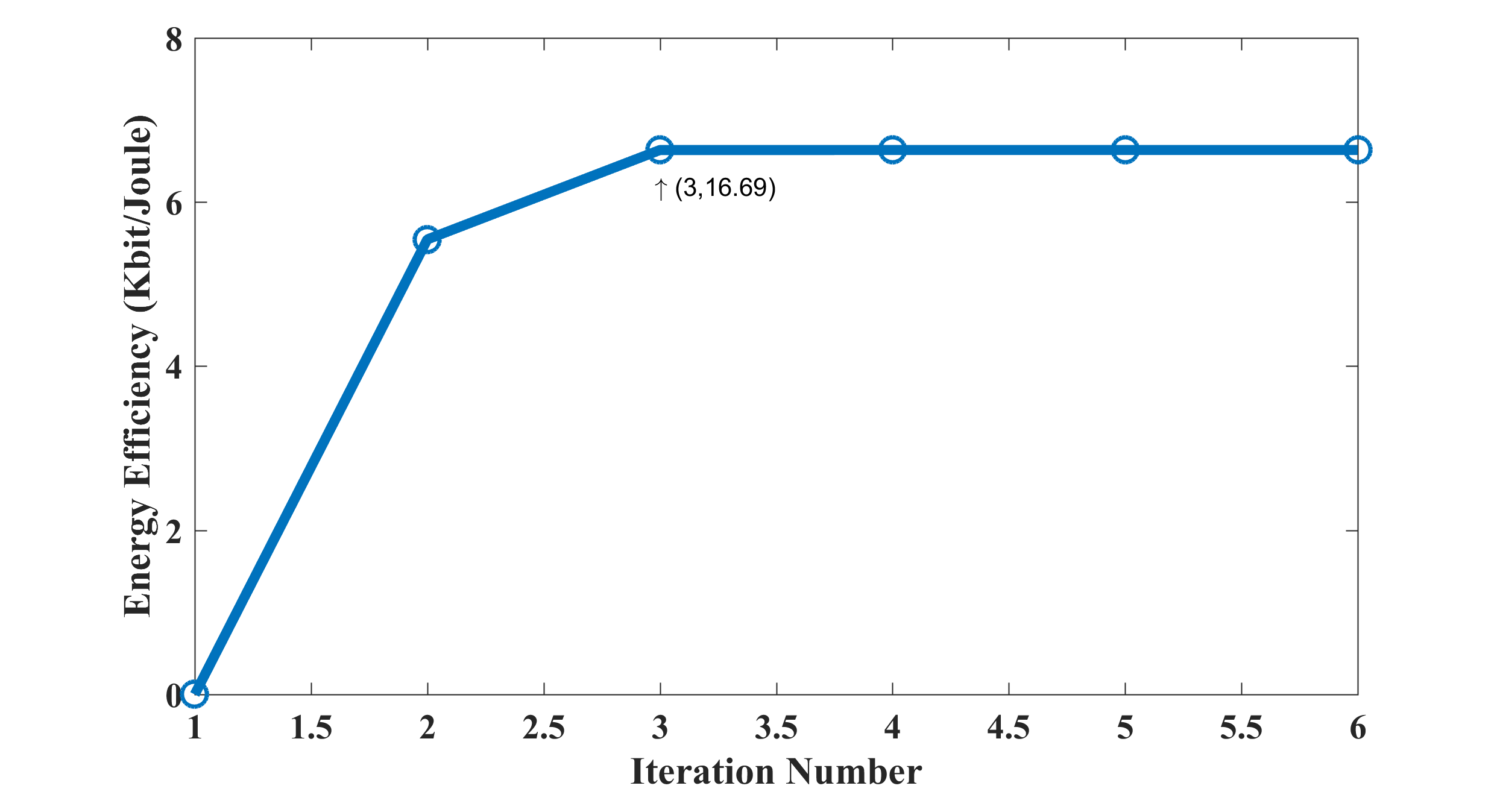}
\caption {Convergence of the proposed algorithm for energy efficiency at $M = 100$}
\end{figure}
\begin{figure}
\center
\includegraphics[scale = 0.51]{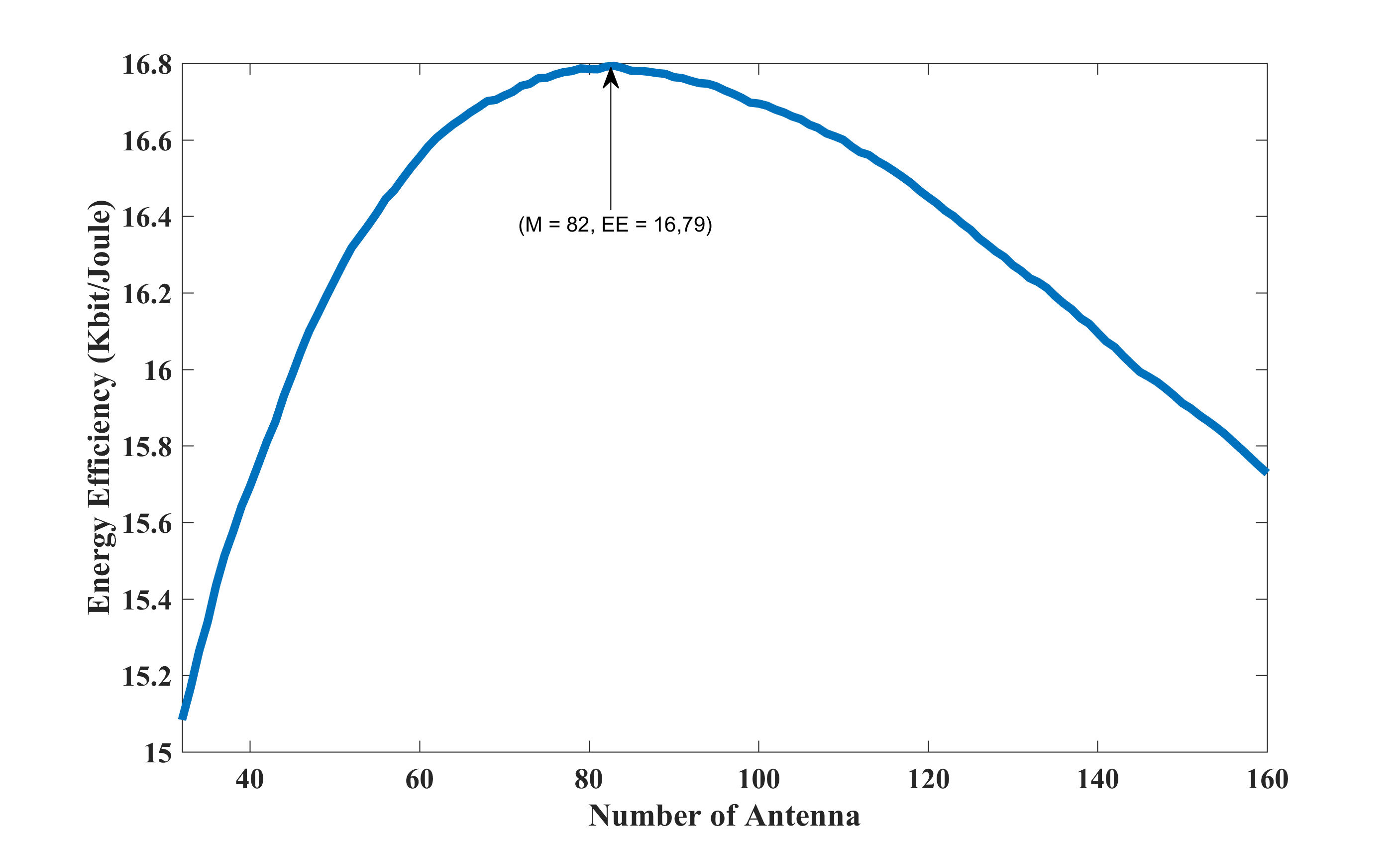}
\caption {Energy efficiency versus number of transmit antenna for $K = 5$}
\end{figure}
\begin{figure}
\center
\includegraphics[scale = 0.51]{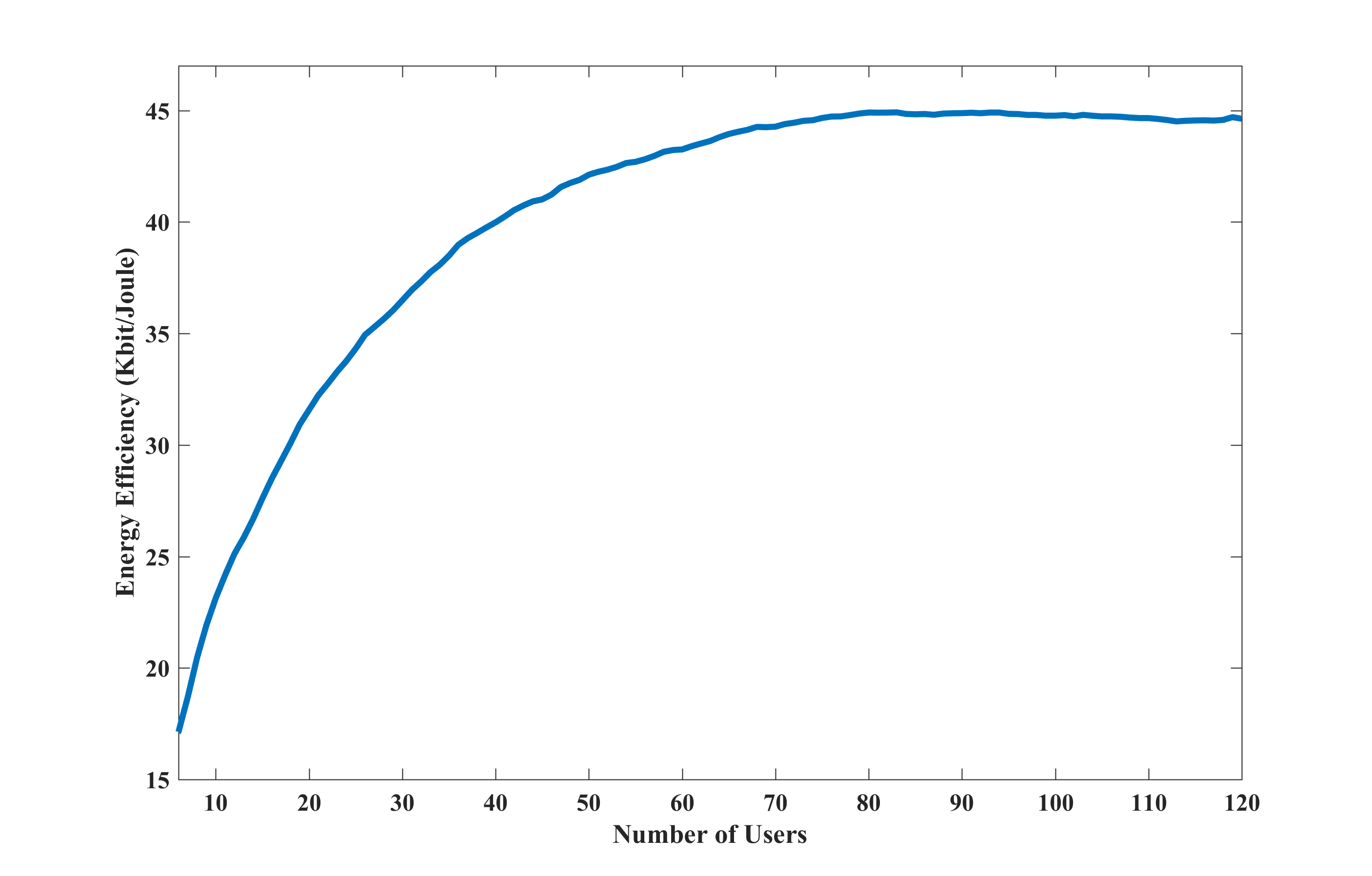}
\caption{Energy efficiency versus number of users for $M = 82$}
\end{figure}
\begin{figure}
\center
\includegraphics[scale = 0.51]{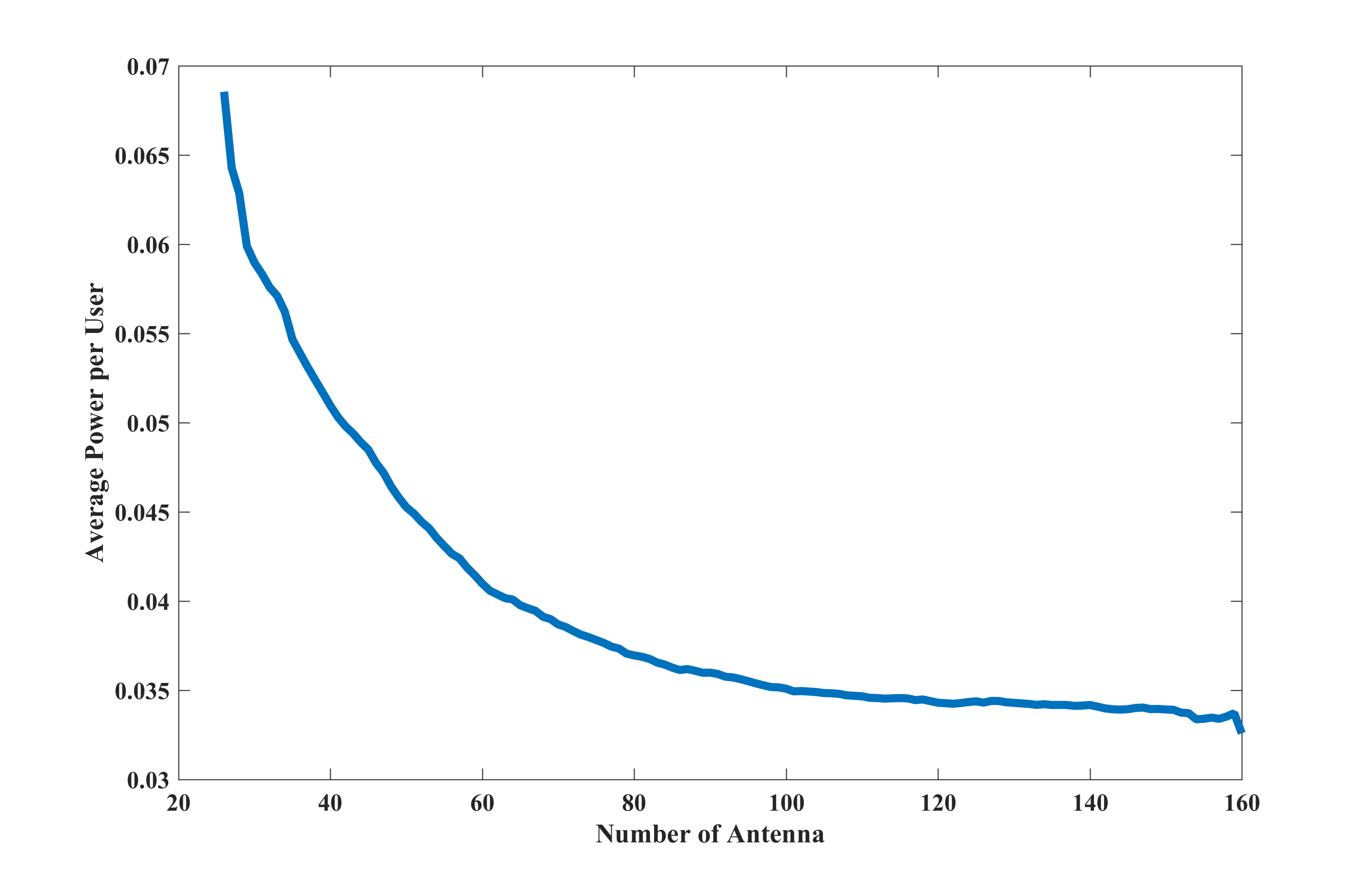}
\caption{Average transmit power per one user over number of transmit antenna}
\end{figure}
\begin{figure}
\center
\includegraphics[scale = 0.71]{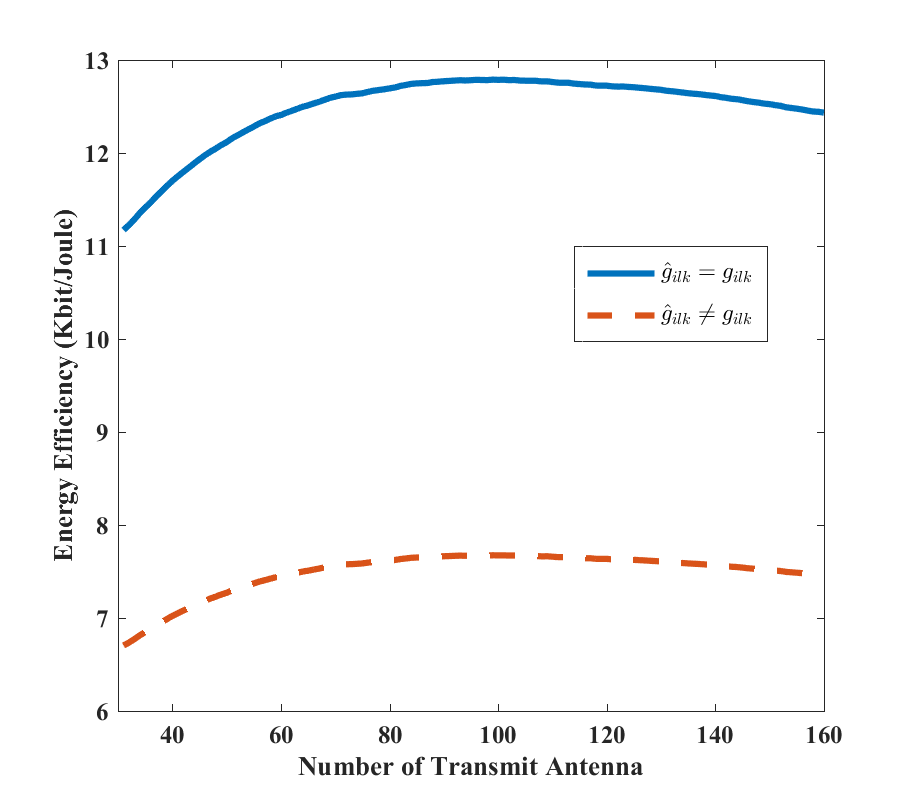}
\caption{Energy efficiency versus number of transmit antenna for perfect and imperfect CSI}
\end{figure}
Fig. 6 shows energy efficiency in multi cell versus transmit antenna. Transmit power is obtained by Algorithm 1 when the estimated channel is in access, by utilizing designed transmit power, a bit with BPSK modulation in a real channel is transmitted. Energy efficiency for any number of transmit antenna in receiver is computed, this process is implemented 10000 times for any number of antenna whenever the channel changes, and the average of energy efficiency over 10000 is plotted. Energy efficiency in multi cell related to single cell, due to pilot contamination and inter cell interference, decreased to about 0.75 of single cell. In channel estimation with pilot training, estimated channel is summation of desired user channel and the other users with same pilot sequence in neighbor cells, so MRT beam-forming , which uses channel vector directly, in addition to focus on desire user, also focus an amount of power to all users with the same pilot. By increasing number of transmit antenna interference on users with same pilot in the other cells gets higher. Therefore, in large number of transmit antenna when we use MRT beam-forming, inter-cell interference increases and the energy efficiency is placed in less than 0.75 of single cell.

\section{Conclusion}
In this paper, a cooperative energy efficient power allocation algorithm for downlink massive MIMO system is proposed. Based on cooperation between users and BS, a closed form for optimal transmit power is found and  an algorithm is proposed. Simulation results show convergence of the proposed algorithm. The simulations show that there is an optimal number of antenna and optima number of users in each cell due to circuit power consumption and inter user interference. In addition, the simulations shown that the more number of transmit antenna the less transmit power. Finally the scenario is extended to multi cell case, in which pilot contamination is considered. The network total energy efficiency decreases in comparison to single cell case. Also, results show that considering pilot contamination for large number of transmit antenna with MRT beamforming increases inter cell interference.


%

\appendices
\section{Transmit Power for Multi Cell Case}
Each user sends a pilot with power $p^u$ to its corresponding BS. The received signal at $j$th BS from users, ${\bf Y}_j$, can be written as follows.
\begin{equation}
{\bf Y}_j = \sum_{l = 1}^{L}\sqrt{p^u}{\bf G}_{jl}{\bf{\Phi}}_{l}^H + {\bf Z}_j
\end{equation}
where ${\bf Y}_j$ is a $M\times \tau $ matrix, ${\bf G}_{jl}=[{\bf g}_{jl1}\: ...\:{\bf g}_{jlK} ]$ is a $M\times K$ matrix which shows channel gains between users in cell $l$ and $j$th BS, ${\bf \Phi}_l =[{ \phi}_{jl} \: ... \: { \phi}_{lK}]$ is $K\times \tau$ pilot matrix of users in cell $l$ that ${\bf \Phi}_l \times {\bf \Phi}_l^H = {\bf I_K} $ and ${\bf \Phi}_{l_1} \times {\bf \Phi}_{l_2} \neq 0\:\: \forall \:l_1 \neq l_2 $ due to pilot reuse. By utilizing LS estimation method $\hat {\bf g}_{jlk}$ can be obtained as following
\begin{equation}
\hat {\bf g}_{jlk} = \frac{1}{\sqrt{p^u}}{\bf Y}_j{\phi}_{lk}
\end{equation}

Received signal to interference plus noise ratio ({SINR}) of user $m$ in cell $j$ can be obtained as following.
\begin{equation}
sinr_{jm} = \frac{p_{jm} \|\hat {\bf g}_{jjm} {\bf w}_{jm}  \|^2}{\sum_{l=1}^{L}\sum_{k=1,\neq m}^{K} {p_{lk}} \|\hat{\bf g}_{ljm} {\bf w}_{lk}  \|^2 + \sigma^2}
\end{equation}
Therefore the transmission rate for the same user can be formulated as follows.
\begin{equation}
r_{jm} = B\log_{2}({1+\Gamma sinr_{jm}})
\end{equation}
Finally, the energy efficiency of the network can be expressed as
\begin{equation}
\eta = \frac{ \sum_{l=1}^{L}\sum_{k=1}^{K}r_{lk}}{\sum_{l=1}^{L}\sum_{k=1}^{K}p_{lk}+ \sum_{l=1}^{L}P_l^c}=\frac{A({\bf P})}{B({\bf P})}
\end{equation}
where $P_l^c$ is circuit power consumption in cell $l$ and $\bf P$ is $L\times K$  transmit power matrix, the energy efficiency maximization problem can be expressed as
\begin{equation} \label{eq25-prob}
\begin{array}{rlllll}
\displaystyle {\max_{\bf P} }& \multicolumn{1}{l}{\eta=\frac{A({\bf P})}{B({\bf P})}} \\
\textrm{s.t.} &  C1: \sum_{k=1}^{K}p_{lk}\leq P^{max}\:\: \forall\:\: l\\
&\displaystyle  C2: r_{lk}\geq R^{min}
\end{array}
\end{equation}

 Using Dinkelbach and SCA algorithms the problem \eqref{eq25-prob} can be written
\begin{equation} \label {eq:19}
\begin{array}{rlllll}
\displaystyle {\max_{\hat {\bf{P}}} }& \multicolumn{1}{l}{A({\hat {\bf{P}}})-\eta B({\hat {\bf{P}}})} \\
\textrm{s.t.} &  C1: \sum_{k=1}^{K}e^{\hat{p}_{lk}}\leq P^{max}\:\: \forall\:\: l\\
&\displaystyle  C2: r_{lk}\geq R^{min}
\end{array}
\end{equation}
The problem \eqref{eq:19} is a convex optimization problem and utilizing Lagrange method, transmit power and Lagrange multipliers are updated as follows
\begin{equation} \label{opt.P4}
p_{jm} = [\frac{(\lambda_{jm}+1)\frac{B\alpha_{jm}}{\ln2}}{\frac{B}{\ln2}\sum_{l=1}^{L}\sum_{k=1,k\neq m}^{K}\frac{z_{lk}}{I_{lk}}-(\eta +\phi_j)}]^+
\end{equation}
\begin{equation} \label{eq:26}
\lambda_{lk}(t_3+1) =(\lambda_{lk}(t_3)-\gamma_\lambda (r_{lk}^{min}-R^{min}))^+ 
\end{equation}
\begin{equation}\label{eq:27}
\phi_{l}(t_3+1) =(\phi_{l}(t_3)-\gamma_\phi (P^{max}-\sum_{k = 1}^{K}p_{lk}))^+ 
\end{equation}
where $\gamma_\lambda$ and $\gamma_\phi$ are step size and $t_3$ is Lagrange iteration index.


\begin{thebibliography}{99}

\bibitem{survey}
M. U. Farooq, M. Waseem, M. T. Qadri and M. Waqar, \textquotedblleft Understanding 5G wireless cellular network: challenges, emerging research directions and enabling technologies,\textquotedblright \textit{ Wireless Personal Communications}, 95, no. 2, pp. 261-285, Jul. 2017.


\bibitem{survey2}
C. Sexton, N. Kaminski, J. M. Marquez-Barja, N. Marchetti and L. A. DaSilva, \textquotedblleft 5G: Adaptable networks enabled by versatile Radio Access Technologies,\textquotedblright \textit{ IEEE Communications Society}, vol. 19, no. 2, pp. 688-720, Jan. 2017.

\bibitem{massive}
Z. Bojkovic and D. Milovanovic, \textquotedblleft{A technology vision of the fifth generation (5g) wireless mobile networks},\textquotedblright \textit{ International Conference on Emerging Trends in Electrical, Electronic and Communications Engineering}, Springer, pp.25-43, Cham, Nov. 2016.

\bibitem{massive2}
E. Björnson, J. Hoydis, L. Sanguinetti, \textquotedblleft{Massive MIMO networks: spectral, energy, and hardware efficiency},\textquotedblright \textit{ Foundations and Trends in Signal Processing}, vol.11, no.3-4, pp.154–655, 2017.[Online].Available:http://dx.doi.org/10.1561/2000000093



\bibitem{CP}
L. Zhao, K. Li, K. Zheng, and M. O. Ahmad, \textquotedblleft An analysis of the trade off
between the energy and spectral efficiency in an uplink massive MIMO-OFDM system,\textquotedblright \textit{ IEEE Trans. Circuits Systems II, Exp. Briefs}, vol. 62, no. 3, pp. 291-295, Mar. 2015.

\bibitem{ee41}
C. Li, Y. Li, K. Song and L. Yang, \textquotedblleft Energy efficient design for multiuser downlink energy and uplink information transfer in 5G,\textquotedblright\textit{ Science China Information Sciences}, 59, no. 2, Feb. 2016.

\bibitem{ee42}
K. Guo, Y. Guo, and G. Ascheid, \textquotedblleft Energy-Efficient Uplink Power Allocation in
Multi-Cell MU-Massive-MIMO Systems,\textquotedblright\textit{ Proceedings of 21th European Wireless Conference}, 2015.

\bibitem{ee43}
Y. Li, C. Tao, L. Liu, L. Zhang, \textquotedblleft Energy-efficiency-aware relay selection in
distributed full duplex relay network with massive MIMO,\textquotedblright\textit{ Science China Information Sciences}, 60, no. 2,  Feb. 2017.


\bibitem{ee51}
L. Zhao, H. Zhao, F. Hu, K. Zheng, and J. Zhang, \textquotedblleft Energy efficient power
allocation algorithm for downlink massive MIMO with MRT precoding,\textquotedblright\textit{ In Vehicular Technology Conference (VTC Fall), 2013 IEEE 78th}, pp. 1–5, 2013.

\bibitem{ee52}
J. Arshad, J. Li,T. Younas, M. Sheng and L. Hongyan, \textquotedblleft Analysis of energy efficiency and area throughput in large scale MIMO systems with MRT and ZF precoding,\textquotedblright\textit{ Wireless Personal Communications}, 96, no. 1, pp. 23-46, Sep. 2017.

\bibitem{ee6}
E. Björnson, L. Sanguinetti, J. Hoydis, M. Debbah, \textquotedblleft Optimal Design of Energy-Efficient Multi-User MIMO Systems: Is Massive MIMO the Answer?,\textquotedblright\textit{ IEEE Transactions on wireless Communication}, vol. 14, no. 6, pp. 3059-3075, June 2015.

\bibitem{ee71}
Y. GUO, J. TANG, G. WU and S. LI, \textquotedblleft Power allocation for massive MIMO: impact of power amplifier efficiency,\textquotedblright\textit{ Science China Information Sciences}, 59, no. 2,  Feb. 2016.

\bibitem{ee72}
Z. Liu, W. Du, and D. Sun, \textquotedblleft Energy and spectral efficiency tradeoff for
massive MIMO systems with transmit antenna selection,\textquotedblright\textit{ IEEE Trans.
Vehi. Tech.}, vol. 66, no. 5, pp. 4453-4457, May 2017.

\bibitem{ee8}
T. X. Tran and K. C. Teh, \textquotedblleft Energy and spectral efficiency of leakagebased
precoding for large-scale MU-MIMO systems,\textquotedblright\textit{ IEEE Commun.
Lett.}, vol. 19, no. 11, pp. 2041–2044, Nov. 2015.

\bibitem{ee9}
Y. Zhou, D. Li, H. Wang, A. Yang, and S. Guo, \textquotedblleft QoS-aware energy-efficient optimization for massive MIMO systems in 5G,\textquotedblright\textit{ In Sixth International Conference on Wireless Communications and Signal Processing}, pages 1–5, 2014.

\bibitem{ee10}
T. M. Nguyen and L. B. Le, \textquotedblleft Joint pilot assignment and resource allocation in multicell massive MIMO network: Throughput and energy efficiency maximization,\textquotedblright\textit{ In 2015 IEEE Wireless Communications and Networking Conference (WCNC)}, pp. 393–398, 2015.

\bibitem{pcon}
F. Rusek, D. Persson, B. K. Lau, E. G. Larsson, T. L. Marzetta, O. Edfors, and F. Tufvesson, \textquotedblleft Scaling up MIMO: Opportunities and challenges with very large arrays,\textquotedblright\textit{ IEEE Signal Process. Mag.}, vol. 30, pp. 40–60, Jan. 2013.

\bibitem{shgap}
A. Goldsmith, \textit{Wireless Communications.} Cambridge, U.K.: Cambridge
Univ. Press, 2005.

\bibitem{ee1}
H, Hamdoun, P. Loskot, T. O’Farrell and J. He, \textquotedblleft Survey and applications of standardized energy metrics to mobile networks,\textquotedblright\textit{ Annals of telecommunications}, 67, no. 3-4, pp. 113-123, Apr. 2012.

\bibitem{ee2}
Y. Chen, S. Zhang, S. Xu, and G. Li, \textquotedblleft Fundamental trade-offs on green
wireless networks,\textquotedblright\textit{ IEEE Commun. Mag.}, vol. 49, no. 6, pp. 30–37,
2011.

\bibitem{dinkel}
W. Dinkelbach, \textquotedblleft On nonlinear fractional programming,\textquotedblright\textit{Bulletin of the
Australian Mathematical Society}, vol. 13, pp. 492–498, Mar. 1967.

\bibitem{scale}
H. Zhang, Y. Fang, Y. Xie, H. Wu and Y. Guo, \textquotedblleft Power control for multipoint cooperative
communication with high-to-low sinr scenario,\textquotedblright\textit{ EURASIP Journal on Wireless Communications and Networking}, 2012, no. 1, pp. 276, Dec. 2012.

\bibitem{boid}
S. Boyd and L. Vandenberghe, \textit{Convex Optimization.} Cambridge,
U.K.: Cambridge Univ. Press, 2004.

\bibitem{ls}
S. M. Kay, \textit{Fundamentals of Statistical Signal Processing, Volume I:
Estimation Theory. Pearson Education}, vol. 1, 1993.







\end{thebibliography}
\end{document}